%% file: paper.tex
\documentclass[aps,prd,reprint,superscriptaddress,floatfix,showpacs]{revtex4-2}
\usepackage{graphicx}
\usepackage[capitalise]{cleveref}
\usepackage[english]{babel}
\usepackage{placeins}
\usepackage{longtable}
\usepackage{dcolumn}
\usepackage{xcolor}
\usepackage{siunitx}

\usepackage{lineno}

\usepackage{ulem}
\DeclareSIUnit\clight{\ensuremath{c}}
\newcommand{\KS}{\ensuremath{K_S^0}}
\newcommand{\JPsi}{\ensuremath{J\!/\!\psi}}
\newcommand{\Sigp}{\ensuremath{{\Sigma}^+}}
\newcommand{\pbar}{\ensuremath{\bar p}}
\newcommand{\p}{\ensuremath{p}}
\newcommand{\Sigpbar}{\ensuremath{\bar{\Sigma}^-}}
\newcommand{\SigDecay}{\ensuremath{\JPsi\to\pbar\Sigp\KS}}
\newcommand{\SigbarDecay}{\ensuremath{\JPsi\to p\Sigpbar\KS}}
\newcommand{\pip}{\pi^+}
\newcommand{\pim}{\pi^-}

\newcommand{\piz}{\pi^0}
\newcommand{\BF}{\ensuremath{\mathcal{B}} }

\begin{document}
\input{Results}

\title{\boldmath Observation and branching fraction measurement of the decay $\SigDecay+c.c.$}

\include{AuthorList}

\begin{abstract}
    The first observation of the decays $\SigDecay$ and $\SigbarDecay$ is reported using $(10087\pm44)\times10^{6}$ $\JPsi$ events recorded by the {BESIII} detector at the {BEPCII} storage ring. The branching fractions of each channel are determined to be $\BF(\SigDecay)=\BFresultstsySigma$ and $\BF(\SigbarDecay)=\BFresultstsySigmabar$. The combined result is $\BF(\SigDecay+c.c.)=\BFresultstsy$, where the first uncertainty is  statistical and the second systematic. The results presented are in good agreement with the branching fractions of the isospin partner decay $\JPsi\to pK^-\bar\Sigma^0+c.c.$. 
\end{abstract}

\date{\today}


\maketitle

\section{Introduction}

The Standard Model of particle physics describes most aspects of nature with very high precision. However, there are still many topics left where the experimental observations are not understood in detail.
Especially in the non-perturbative regime of quantum chromodynamics (QCD), it is difficult to obtain accurate predictions for particle interactions, resonance spectra and decay processes. For example, the spectrum of excited nucleon states ($N^*$ resonances) is still not fully understood. Although a large number of $N^*$ states are predicted by theoretical approaches, only a subset of these has been confirmed by experiments to date. The majority of the observed states, as listed by the Particle Data Group (PDG)~\cite{pdg}, are poorly understood and reported only by one experiment. Often they are only observed in decays to non-strange final states. To determine the internal structure of these resonances,
it is also necessary to investigate possible decays of the $N^*$ resonances into final states with strange-quark content, {\it e.g.} hyperons and kaons.

$N^*$ resonances with intrinsic strangeness are accessible by the decay channels $\SigDecay$ and $\SigbarDecay$, in which their properties can be investigated.  A large branching fraction of the $N^*$ resonances to these final states indicates an intrinsic strangeness content already present in the respective resonance. In addition, excited $\Sigma$ states are also produced and can be investigated in these decays. 
In this paper, the first observation of the decay channel $\SigDecay$ and its charge conjugate $\SigbarDecay$ together with the first determination of the single decay branching fractions are presented together with the combined branching fraction $\BF(\SigDecay+c.c.)$.
The branching fraction of the isospin partner $\JPsi\to pK^-\bar\Sigma^0+c.c.$ was measured to be 
\begin{equation}
    \BF(\JPsi\to pK^-\bar\Sigma^0+c.c.)=\BFpkplusSigma
\end{equation}
by the Mark II experiment at the SPEAR accelerator using $\num{1.32}\times 10^6$ $\JPsi$ events. This indicates that the branching fraction of the decay of interest for this paper should have a similar order of magnitude. Therefore, given the dataset at BESIII where 10 billion $\JPsi$ events have been recorded and improvements in the analysis method were made, especially for the determination of systematic uncertainties, 
a high precision measurement of the decay channel $\SigDecay+c.c.$ is possible.

\section{BESIII experiment}
The BESIII detector is a magnetic
spectrometer~\cite{BESIIISYS:2009} located at the Beijing Electron
Positron Collider (BEPCII)~\cite{Yu:IPAC2016TUYA01}. The
cylindrical core of the BESIII detector consists of a helium-based
multilayer drift chamber (MDC), a plastic scintillator time-of-flight
system (TOF), and a CsI(Tl) electromagnetic calorimeter (EMC),
which are all enclosed in a superconducting solenoidal magnet
providing a 1.0~T magnetic field \cite{detvis}. The solenoid is supported by an
octagonal flux-return yoke with resistive plate chamber muon-identifier
modules interleaved with steel. The acceptance for
charged particles and photons is 93\% over the $4\pi$ solid angle. The
charged-particle momentum resolution at $1~\si{\giga\eV/\clight}$ is
$0.5\%$, and the the specific ionization energy loss $\mathrm{d}E/\mathrm{d}x$ resolution is $6\%$ for electrons
from Bhabha scattering. The EMC measures photon energies with a
resolution of $2.5\%$ ($5\%$) at $1$~GeV in the barrel (end-cap)
region. The time resolution of the TOF barrel part is 68~ps.
The time resolution of the end-cap TOF
system was upgraded in 2015 with multi-gap resistive plate chamber
technology, providing a time resolution of
60~ps~\cite{etof::Guo2019,etof::Li2017}; this upgrade benefits about 87\% of the total dataset analyzed here.

\section{\label{sec::DATASETS}Data sets and Monte Carlo simulation}

For the determination of the branching fraction of the decay channel $\SigDecay$ (here in the following charge conjugation is implied), the complete $\JPsi$ data sample recorded in the years 2009, 2012, 2018 and 2019 by the BESIII experiment is analyzed. The total number of $\JPsi$ events is determined by using inclusive $\JPsi$ decays with the method described in Ref.~\cite{BESIII:2012pbg}.
To correct for $\JPsi$ candidates that originate from background contributions due to QED processes, beam-gas interactions, and cosmic rays, continuum data samples recorded at $\sqrt{s} = \SI{3.080}{\giga\eV}$ are used.
The detection efficiency for the inclusive $\JPsi$ decays is obtained using the experimental data sample of $\psi(3686) \rightarrow \pi^+\pi^-\JPsi$.
The efficiency difference between the $\JPsi$ produced at rest and the $\JPsi$ from the decay $\psi(3686)\rightarrow\pi^+\pi^-\JPsi$ is estimated by comparing the corresponding efficiencies in Monte Carlo (MC) simulation.
The uncertainties related to the MC model, track reconstruction efficiency, fit to the $\JPsi$ mass peak, background estimation, noise mixing, and reconstruction efficiency for the pions recoiling against the $\JPsi$ are studied.
Finally, the number of $\JPsi$ events is determined to be $N_{\JPsi}= (10087\pm44)\times10^{6}$, where the uncertainty includes both statistical and systematic uncertainties \cite{BESIII:2012pbg}.

For the optimization of the analysis procedure and the determination of the reconstruction efficiency, MC samples are generated. The initial collision is handled by KKMC \cite{Jadach:2000ir} to take into account initial state radiation. Subsequently,
the reaction particles are generated with the event generator {\sc evtgen} \cite{ref::evtgen2001,ref::evtgen2008} using world-average branching fractions. The following interaction with the detector and further decays of the primary particles are simulated with the {\sc geant4} package~\cite{Agostinelli:2002hh}. The
signal MC sample is generated from a phase space distributed sample by using
the results of the amplitude analysis described later in the paper as weights.

For this work two MC samples are used. The first one is needed to calculate the reconstruction efficiency of the signal decay.
In this sample only the decay $\SigDecay$ is simulated, where $\Sigp\to\p\piz,\ \piz\to\gamma\gamma$, and $\KS\to\pip\pim$ are exclusively decaying to these final states.
The angular distributions determined from the reconstructed data are taken into account during the calculation of the reconstruction efficiency.
They are adjusted by performing an amplitude analysis with ComPWA \cite{ComPWA}.
To obtain a precise reconstruction efficiency, $4\times10^{6}$ $\JPsi$ events are simulated.
The second MC sample is an inclusive one with $\JPsi$ decaying to anything where all known decay channels are generated in the known ratios to each other. This sample includes both the production of the $\JPsi$ resonance and the continuum processes. 
It is mainly used to identify potential background contributions and consequently the signal events are filtered to form a pure background MC sample.
The sample is generated to match the number of $\JPsi$ events expected in the corresponding BESIII data set.

\section{\label{sec::PID}Event selection}
The decay $\SigDecay$ is reconstructed using the dominant decay channels of the intermediate resonances.
The $\Sigp$\ is reconstructed with the final state $p\pi^0$ with $\pi^0\to\gamma\gamma$.
For the $\KS$,\ the decay into two charged pions ($\KS\to\pip\pim$) is used.
Therefore, each event must have four charged tracks with a total charge of zero and at least two photons.

Charged tracks are required to be reconstructed inside the MDC acceptance ($|\cos\theta|<0.93$ with $\theta$ being the polar angle with respect to the MDC axis).
Additionally, for one anti-proton or one proton in each event the distance of closest approach to the interaction point is required to be within the cylindrical volume around the interaction point ($xy$) with radius $|V_{xy}|<\SI{1}{\centi\meter}$ and in beam direction ($z$) within $|V_z|<\SI{10}{\centi\meter}$.
For the second proton or anti-proton originating from the $\Sigp$ decay, the nearest distance to the interaction point is not restricted in the $xy$ plane but must be less then $\SI{20}{\centi\meter}$ in the $z$ direction.
This value is chosen to take the lifetime of the $\Sigp$ into account.
Furthermore, particle identification (PID) based on the time of flight and the energy loss information is used to reject the pion and kaon hypotheses for the proton.

The photons from the $\piz$ decay are required to have energies greater than $\SI{50}{\mega\eV}$ if they are detected in the end caps
($\num{0.86}<|\cos\theta|<\num{0.92}$) and greater than $\SI{25}{\mega\eV}$ if they are detected in the barrel part ($|\cos\theta<\num{0.8}|$) of the EMC.
The angle between the photon and the nearest charged track is required to be at least $20^\circ$ to exclude photon candidates produced by split-offs of charged tracks.
Furthermore, it is required that the EMC shower time is within an interval of $\SI{700}{\nano\second}$ after the collision, to suppress electronic noise and showers unrelated to the event.
For the $\piz$ selection, the invariant mass $M_{\gamma\gamma}$ is required to be within $[{80,\,180}]\,\si{\mega\eV/\clight^2}$.

No PID is required for the charged pions from $\KS$ decays; a loose constraint on the distance of closest approach to the interaction point is applied ($|V_z|<\SI{20}{\centi\meter}$).
Due to its long lifetime, the $\KS$ is reconstructed by performing a secondary vertex fit.
For all $\KS$ candidates, the ratio of the measured decay length $L(\KS)$  to its uncertainty $\sigma_L$ has to be $L(\KS)/\sigma_L>2$.

After the initial selection, a kinematic fit is performed which uses the momenta of the pions after the vertex fit and the measured values of all other particles.
The kinematic fit constrains the total four-momentum to the one of the initial $e^+e^-$ system and the masses of the $\Sigp$ and the $\piz$ to their known values~\cite{pdg}.
The mass of the $\KS$ is not constrained, and the spectrum of the invariant  mass $M_{\pip\pim}$ is used to determine the number of signal events.
Due to the high number of noise photons, multiple combinations can be  reconstructed in an event. The kinematic fit converges in $2/3$ of all cases only for one candidate and in $1/3$ of the cases for mainly two candidates.
To get rid of these combinations the smallest $\chi^2_{\rm kin}$ of the kinematic fit is used to determine the best candidate. Only a very loose requirement on $\chi^2_{\rm kin}$ is used to ensure convergence of the kinematic fit.

\section{\label{sec::BKGDATA}Background studies}

 The inclusive MC sample is used to examine the major background contributions.
Since it provides the information on each event, such as the generated reaction, it can be used to identify the channels which survive the event reconstruction described above.
The main background channels contain a $\Lambda$ hyperon or an $\eta$ or $\omega$ meson as intermediate state before decaying to the same final state as the decay channel.

\begin{figure}[!htbp]
    \flushleft
    \includegraphics[width=8.0cm]{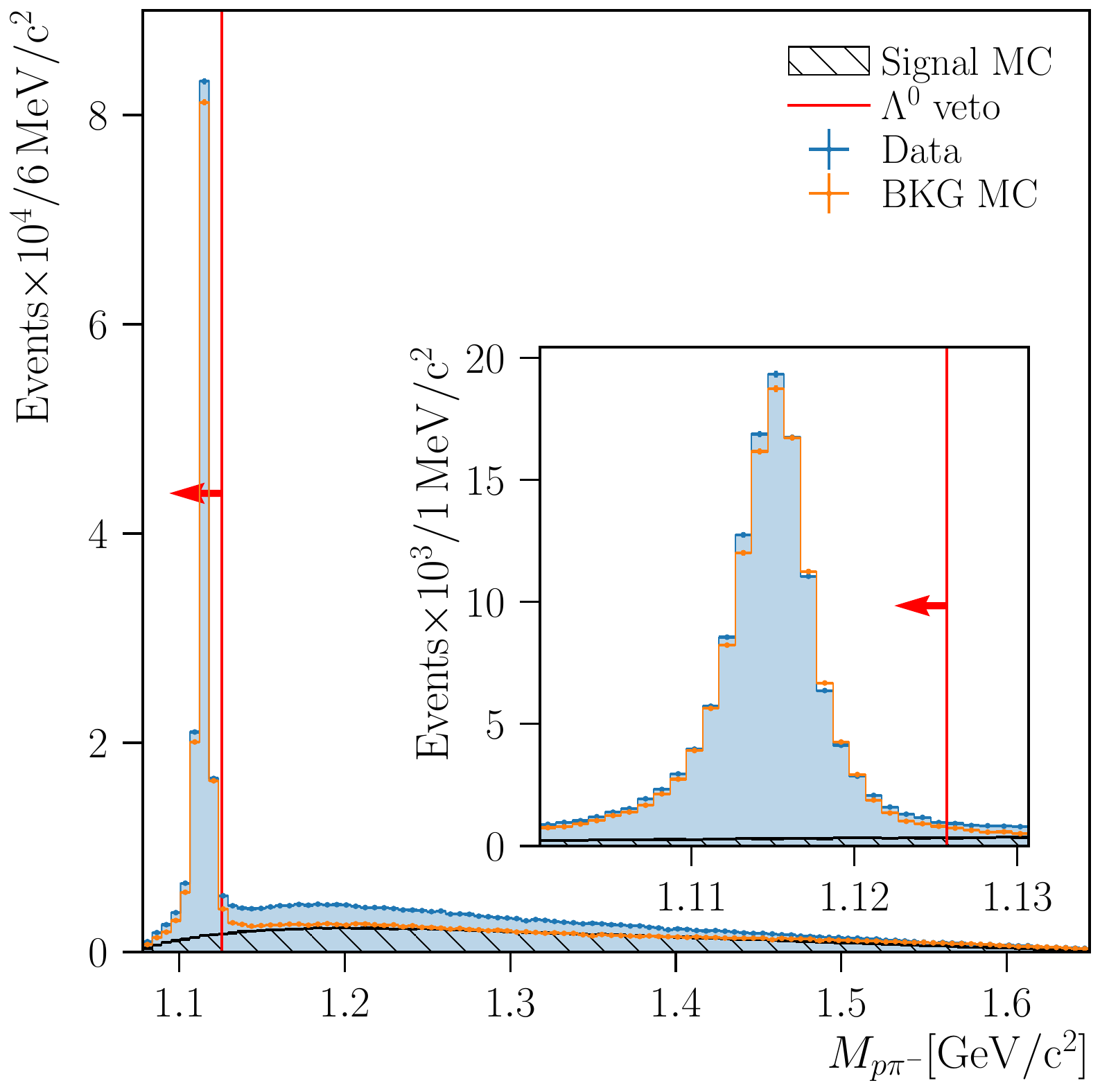}
    \caption{Distribution of $M_{p\pim}$ with the veto indicated to reject decay channels containing a $\Lambda$ as the intermediate state for MC events of the signal channel (black), background events from the inclusive MC sample (orange), and for data (blue).
        The red line indicates the position of the veto and the arrow which events are rejected.
        The inset shows a zoom-in view in the $\Lambda$ peak region.\label{fig:Lambda_veto}}
\end{figure}

Channels which contain a $\Lambda$ hyperon decaying into $\p\pim$, \emph{e.g.} $\JPsi\to\pi^+\Lambda\bar \Sigma^-$, make up the largest contribution to the background. They can be suppressed by rejecting events where the invariant mass $M_{p\pi^-}$ is below $\SI{1.126}{\giga\eV/\clight^2}$.
The requirement is $4\sigma$ above the nominal $\Lambda$ mass, suppressing $98.7\%$ of events with $\Lambda$ in the decay chain.  The requirement causes a signal loss of about $13.5\%$, see \cref{fig:Lambda_veto} for visualization.

\begin{figure}[!htbp]
    \centering
    \includegraphics[width=7.5cm]{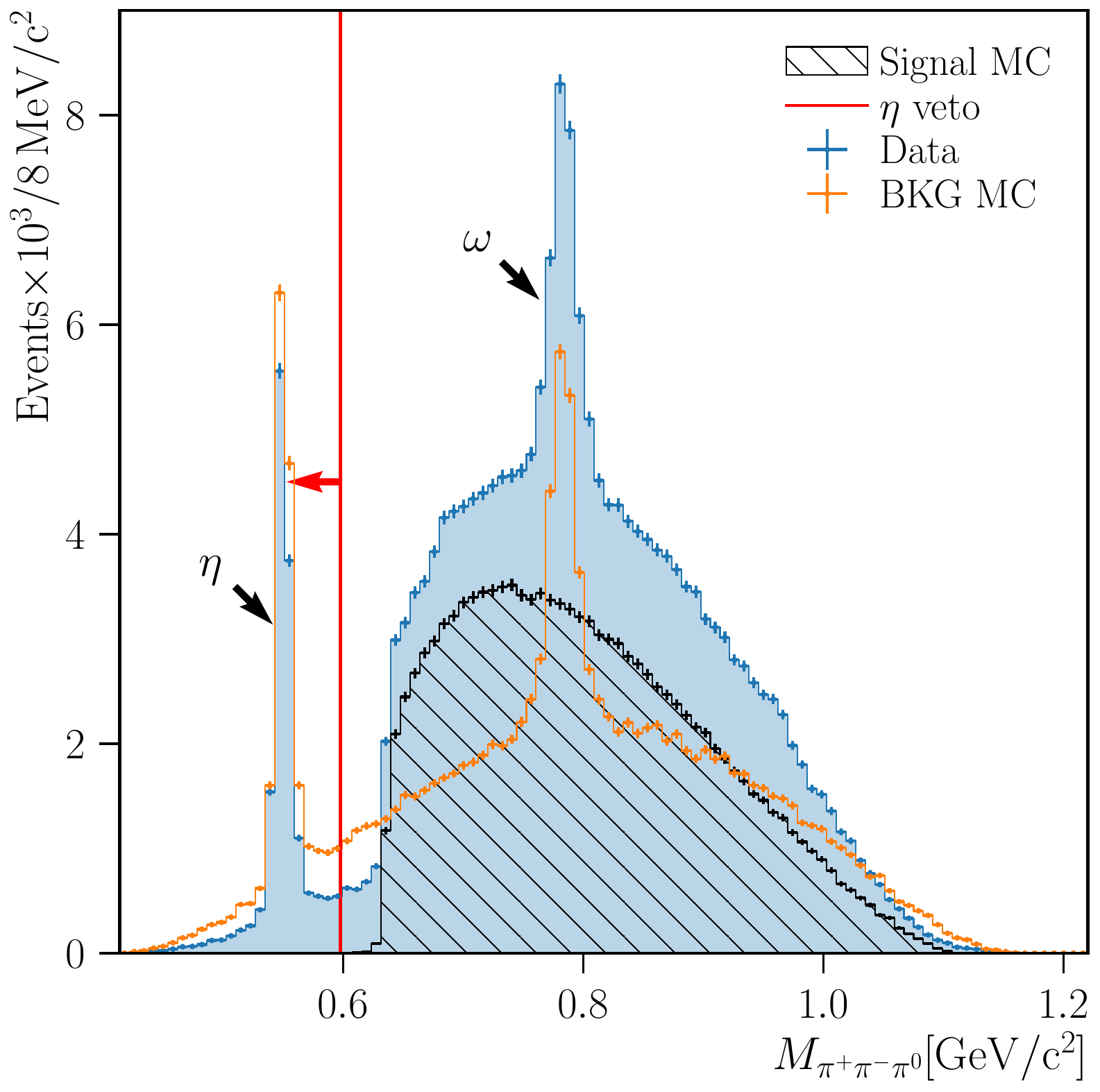}
    \caption{Distribution of $M_{\pip\pim\piz}$ with the veto to reject decay channels containing $\eta$ for MC events of the signal channel (black), background events from the inclusive MC sample (orange), and for data (blue). The red line indicates the position of the veto and the arrow which events are rejected.}
    \label{fig:omega_veto}
\end{figure}

The second largest background contribution stems from the decay $\JPsi\to\p\pbar\eta$ with $\eta\to\pip\pim\piz$, as shown in \cref{fig:omega_veto}.
These events can be easily suppressed, since the $\eta$ mass is below the invariant mass $M_{\pip\pim\piz}$ of the signal decay.
To reject these events a veto on the invariant mass $M_{\pip\pim\piz}$ with $M_{\pip\pim\piz}<\SI{0.598}{\giga\eV/\clight^2}$ is chosen which is $4\sigma$ above the nominal $\eta$ mass and suppresses all $\eta$ related events without any loss of signal events.
The third relevant background channel is $\JPsi\to\p\pbar\omega$ with $\omega\to\pip\pim\piz$, also shown in \cref{fig:omega_veto}.
This decay channel can not easily be suppressed since the $\omega$ peak is sitting in the middle of the distribution of the invariant mass $M_{\pip\pim\piz}$ of the signal events.
Therefore, a large portion of the signal events would be lost by vetoing $\omega$ events.
However, this process does not peak in the $M_{\pip\pim}$ spectrum and can therefore be subtracted statistically.

An additional source of background events is the process $e^+e^-\to\gamma^*\to\pbar\Sigp\KS$ without a $\JPsi$ as intermediate state.
To determine the number of events from the continuum production the data sample recorded at $\sqrt{s}=\SI{3.080}{\giga\eV}$ is analyzed.
The resulting $M_{\pip\pim}$ spectrum is used to estimate the number of background events from continuum production. The yield is $N^{3080}_{\rm QED}=\num{15+-4}$, where the uncertainty
is statistical only. 
Using the $\sqrt{s}=\SI{3.080}{\giga\eV}$ data sample the continuum cross-section can be estimated for the $\sqrt{s}=\SI{3.097}{\giga\eV}$ data set.
Taking the luminosities and the reconstruction efficiencies into account, $N^{3097}_{\rm QED}=\NQED$ QED background events are expected.

\section{\label{sec::BFCalculation} Determination of the Branching Fraction}
The branching fraction $\BF$ of the each signal decay or the combined decays is calculated by
\begin{equation}
    \BF=\frac{N_{\rm Sig}}{N_{\JPsi}}\cdot\frac{1}{\epsilon_{\rm rec}}\cdot\frac{1}{\prod_i{\BF}_i} ,
    \label{eq:BF}
\end{equation}
where $N_{\rm Sig}$ is the number of signal events which is calculated by $N_{\rm Sig}=N_{\KS}-N^{3097}_{\rm QED}$, $N_{\KS}$ is the number of $\KS$, $N_{\JPsi}$ is the number of $\JPsi$ events, $\epsilon_{\rm rec}$ is the reconstruction efficiency, and $\prod_i{\BF}_i$ is the product of the branching fractions of the intermediate states, namely $\BF(\KS\to\pip\pim)$, $\BF(\Sigp\to \p\piz)$, and $\BF(\piz\to\gamma\gamma )$.

\begin{figure}
    \includegraphics[width=7.5cm]{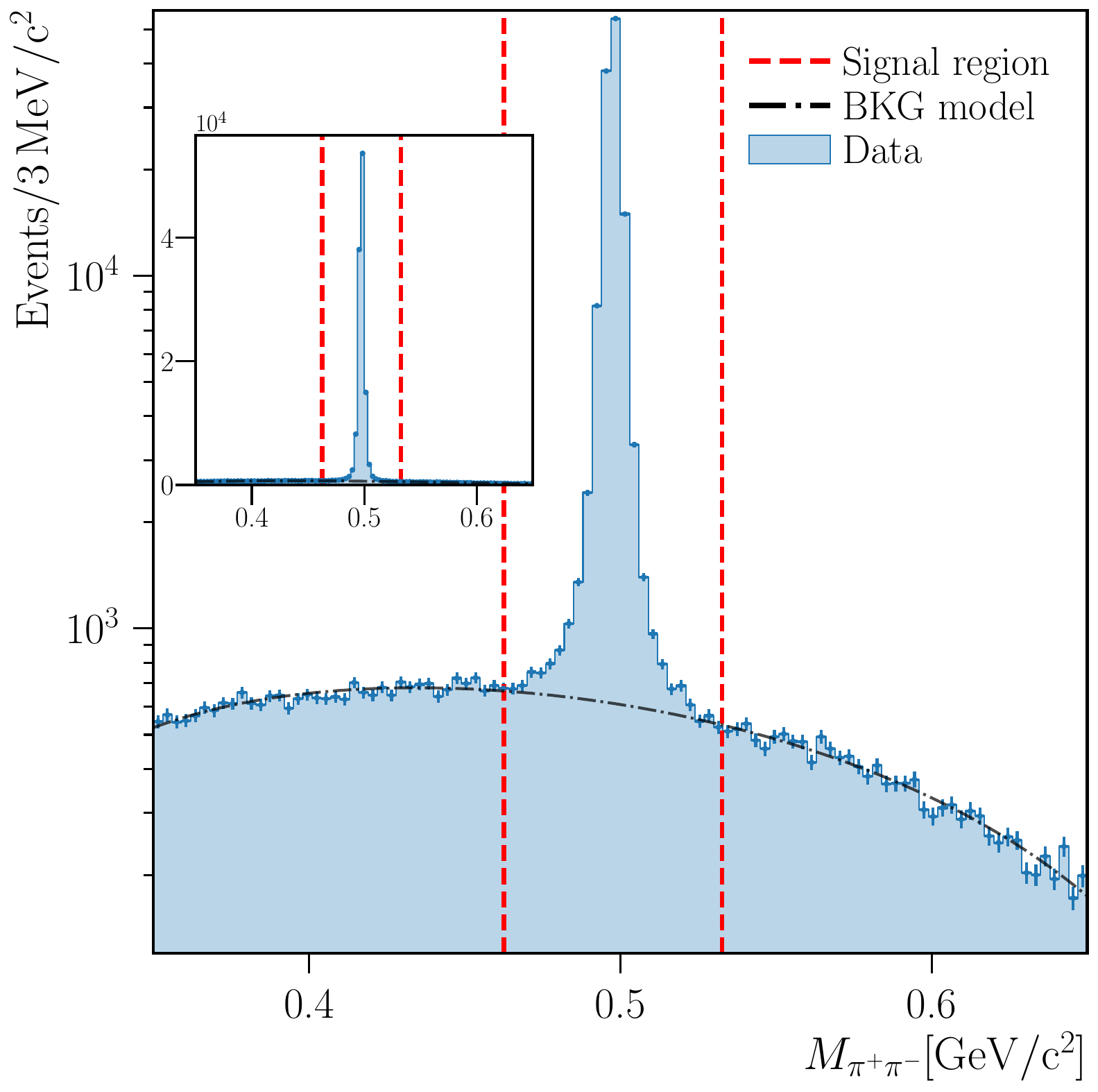}
    \caption{\label{fig:Fit}Distribution of $M_{\pip\pim}$ in data (blue histogram).
        The dashed-dotted black line indicates the background model.
        The dashed red lines show the limits of signal/side band regions. The inset shows the same plot in linear scale.}
\end{figure}

The number of $\KS$ and thus the yield of signal events is determined by counting the number of $\KS$ in the peak of the $M_{\pip\pim}$ distribution above the remaining smooth background contribution (see \cref{fig:Fit}).
For this, as a first step the background shape is obtained by fitting a third-order polynomial to the two side band regions (outside the two dashed red lines), which corresponds to $12\sigma$ of the $K_S$ mass resolution $\sigma$.
Then, this distribution is subtracted from the $M_{\pip\pim}$ distribution and the yield of the remaining entries in the signal region between the two side band regions is determined.
The result is $N_{\KS}=\Nsig$, where the uncertainty is statistical only.

\begin{figure}
    \includegraphics[width=7.5cm]{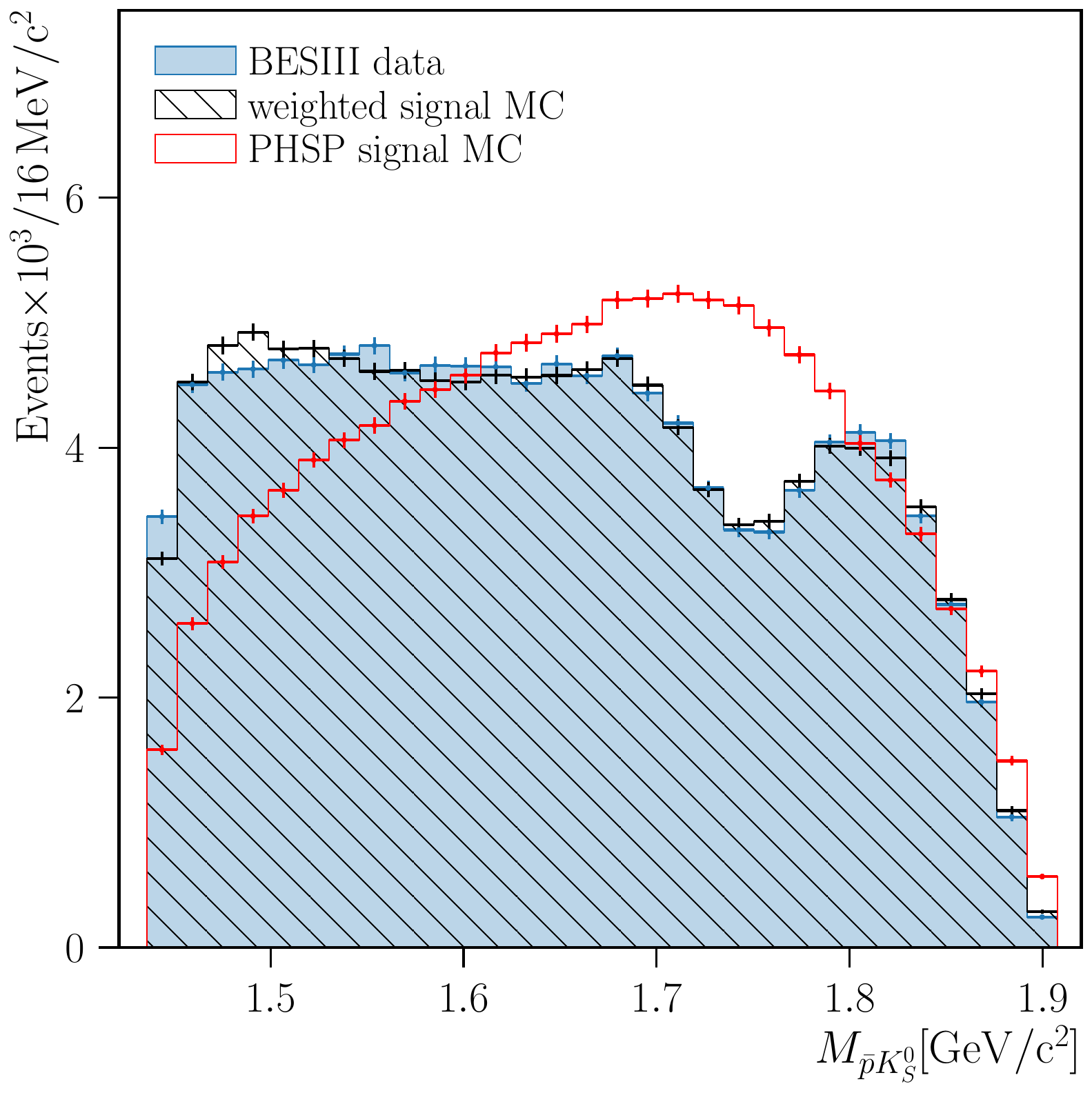}
    \includegraphics[width=7.5cm]{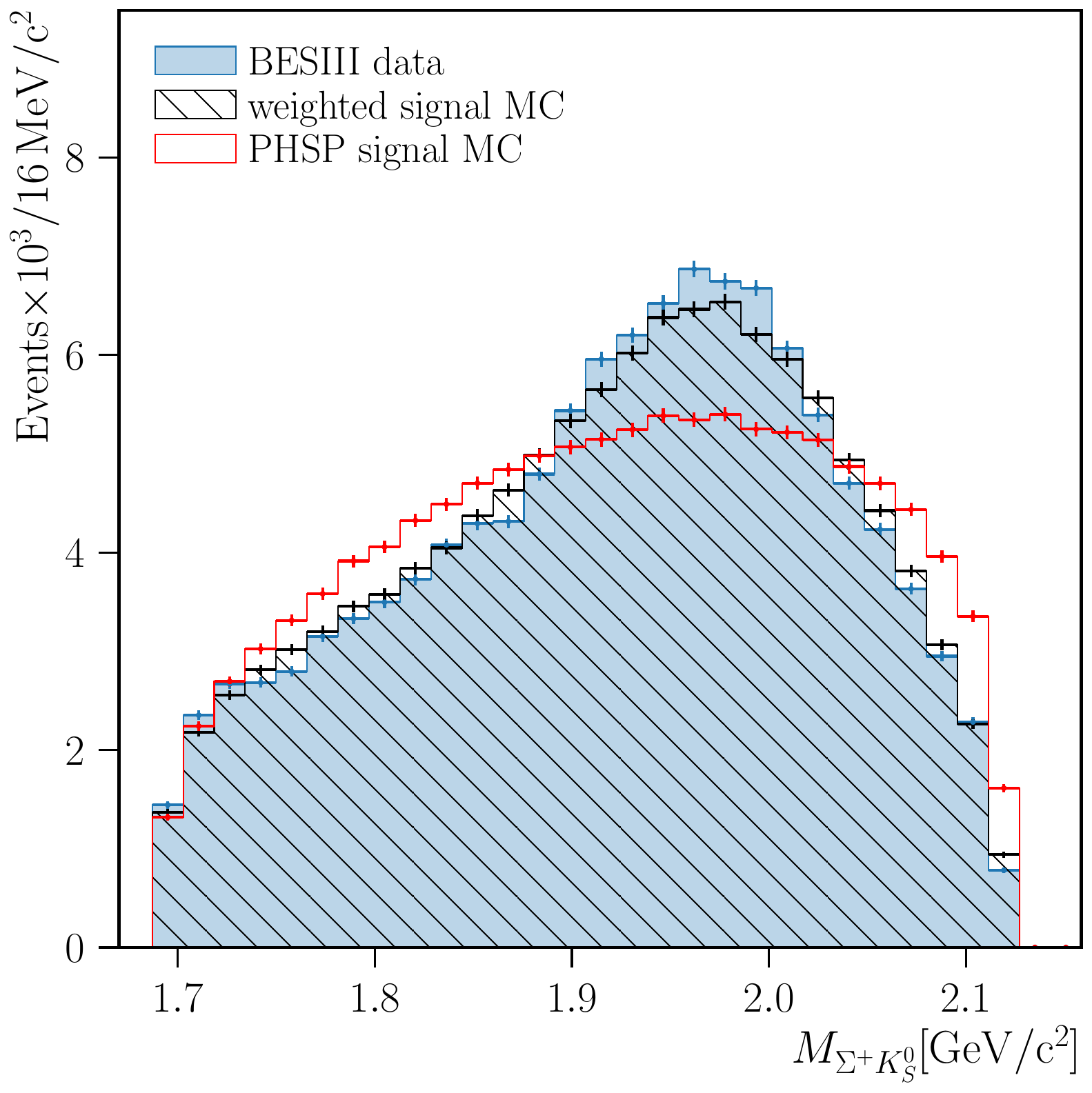}
    \includegraphics[width=7.5cm]{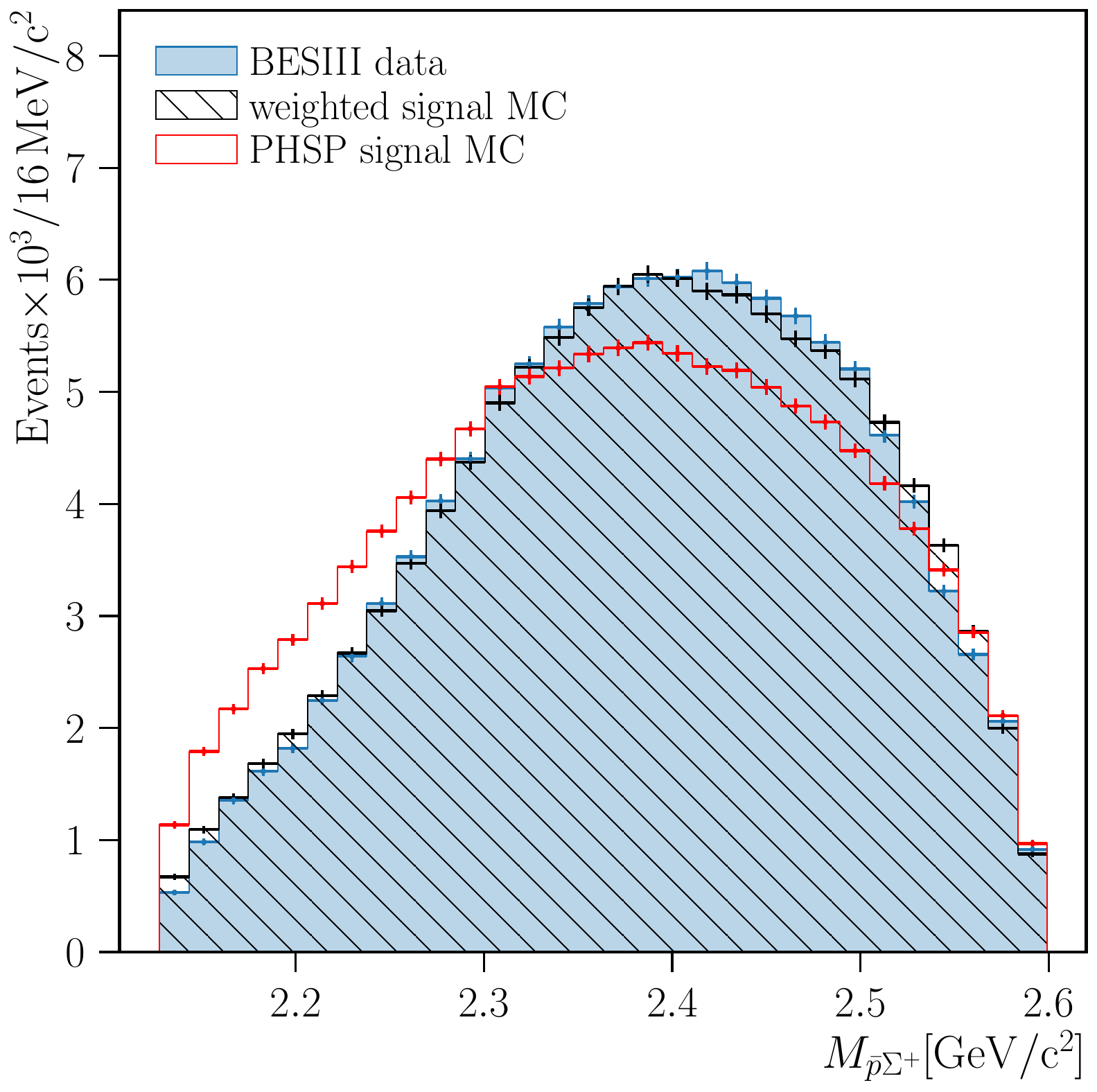}
    \caption{\label{fig:inv_m}Distributions of $M_{\pbar\KS}$, $M_{\Sigp\KS}$, and $M_{\pbar\Sigp}$ from BESIII data (blue histogram).
        The black histogram shows the weighted MC sample, the red histogram the phase space distributed MC sample.       
        }
\end{figure}

The reconstruction efficiency $\epsilon_{\rm rec}$ describes the probability that a signal event is detected and survives the whole selection process.  It depends on the distribution of the final-state particles in the available phase space.
In the analysis of the reaction channel of the isospin partner $\JPsi\to pK^-\bar\Sigma^0+c.c.$ with the MARKII experiment, $\num{90\pm19}$ events were reconstructed. Due to the low number of events, no deviation from the pure phase space distribution was claimed \cite{PhysRevD29804}.
With more than 120\,000 reconstructed events, a large deviation from the three-particle phase space distribution is observed in this analysis (see Fig.~\ref{fig:inv_m}). Therefore, for the determination of the reconstruction efficiency the MC sample is adjusted by using the method of amplitude analysis to match the angular distribution of the BESIII data after event reconstruction. To illustrate this, the distributions of the three invariant masses ($\pbar\KS$, $\Sigp\KS$ and $\pbar\Sigp$) are shown in Fig.~\ref{fig:inv_m}. For all subsystems, the large deviation from the three-particle phase space distribution is corrected for by using an amplitude model with several $\Sigma^*$ and $N^*$ intermediate states. The focus of the adjustment was to properly describe the density of events in the available phase space and thus correctly determine the efficiency.

The reconstruction efficiency is calculated by
\begin{equation}
    \epsilon_{\rm rec}=\frac{N_{\rm rec}}{N_{\rm gen}},
\end{equation}
yielding 
$\epsilon_{\rm rec}~=~\effrec$, where the uncertainty is purely statistical, resulting from limited MC statistics.

With these numbers the combined branching fraction of the decay channels $\SigDecay$ and $\SigbarDecay$ is determined from
\cref{eq:BF} to be
$$
    \BF(\SigDecay+c.c.)=\BFresultst.
$$
The uncertainty is statistical only, taking only the uncertainty of the number of signal events into account.
\cref{tab:calcParameter} shows all relevant parameters.

\begin{table}[h!]
    \centering
    \caption{The parameters used for the determination of the branching fraction.}
    \label{tab:calcParameter}

    \begin{tabular}{lrl}
        \hline\hline
        Parameter                             & \multicolumn{1}{r}{Value}
        \\\hline
        $N_{\JPsi}$ \cite{BESIII:2012pbg}     & $(\num{10086.6+-43.7})$       &
        \hspace{-0.1cm}$\times10^6$
        \\
        $\BF(\KS\to\pip\pim)$ \cite{pdg}      & $(\num{69.20+-0.05})$         & \hspace{-0.1cm}\%
        \\
        $\BF(\Sigp\to p\piz)$ \cite{pdg}      & $(\num{51.57+-0.30}$)         & \hspace{-0.1cm}\%
        \\
        $\BF(\piz\to\gamma\gamma)$ \cite{pdg} & $(\num{98.823+-0.034}$)       &
        \hspace{-0.1cm}\%
        \\\hline
        $N_{\rm Sig}$                         & $(\Nsigvaluecorrected)$       &
        \hspace{-0.1cm}$\times10^5$
        \\
        $N_{\rm QED}^{3097}$                  & \NQED
        \\
        $\epsilon_{\rm rec}$                  & $(\effrecvalue)$              & \hspace{-0.1cm}\%
         \\\hline \hline
    \end{tabular}
\end{table}

In addition, the analysis is performed for each decay channel separately. The results with statistical uncertainties are: 
\begin{eqnarray*}
\BF(\SigDecay)=\BFresultSigma\\
\BF(\SigbarDecay)=\BFresultSigmabar
\end{eqnarray*}

\section{\label{sec::SYSTEMATICS} Systematic uncertainty estimation}

In the following, the different sources of systematic uncertainties are explained. All of them which are determined for the combined result are explained in the following and listed in \cref{tab:sys}.

\begin{table} [b]
    \centering
    \caption{
        The systematic uncertainties of the branching fraction measurement.
        If no value is given, the systematic uncertainty is already covered by the statistical uncertainty.}
    \begin{tabular}{lr}
        \hline\hline
        Source                                             & Uncertainty                           \\\hline\hline
        \multicolumn{2}{c}{Event selection}                                                        \\\hline
        2 (anti-)proton tracks                             & \ppbartracksys                        \\
        2 photon                                           & \totalphotonsys                       \\
        2 PID                                              & \ppbarpidsys                          \\
        $K_S^0$ reconstruction                             & \KSrecsys                             \\
        $\Delta \alpha$                                    & \danglesys                            \\\hline
        \multicolumn{2}{c}{ Background suppression}                                                \\\hline
        $M_{\bar p\pi^+}>\SI{1.126}{\giga\eV/\clight^2}$   & \multicolumn{1}{c}{\hspace{0.9cm}---} \\
        $M_{\pi^+\pi^-\pi^0}>\SI{598}{\mega\eV/\clight^2}$ & \etavetosys                           \\
        kinematic fit                                      & \kinsys                               \\
        $N_{3097}^{\rm QED}$                               & \QEDsys                               \\\hline
        \multicolumn{2}{c}{$M_{\pi^+\pi^-}$ fit}                                                   \\\hline
        Fit range                                          & \FitRangesys                          \\
        Signal range                                       & \multicolumn{1}{c}{\hspace{0.9cm}---} \\
        Background model                                   & \BKGmodelsys                          \\\hline
        \multicolumn{2}{c}{Efficiency}                                                             \\\hline
        Signal MC model                                    & \effmodelsys                          \\
        Signal MC sample size                              & \effstatsys                           \\\hline
        \multicolumn{2}{c}{External}                                                               \\\hline
        $N_{\JPsi}$                                        & \num{0.43}\%                          \\
        $\mathcal{B}(K_S^0\to\pi^+\pi^-)$                  & \num{0.07}\%                          \\
        $\mathcal{B}(\Sigma^+\to p\pi^0)$                  & \num{0.58}\%                          \\
        $\mathcal{B}(\pi^0\to\gamma\gamma)$                & \num{0.03}\%                          \\\hline\hline
        Total                                              & \totalsys                             \\\hline\hline
    \end{tabular}
    \label{tab:sys}
\end{table}

For the determination of the systematic uncertainties concerning the event selection, the differences between data and MC simulated events are evaluated.
For the track reconstruction, a weighting method is used which takes into account the difference in dependence on the transverse momentum and the $\cos\theta$ of the track.
The weights are determined by studying the decay $\JPsi\to \pi^+\pi^- \p \pbar$.
For protons and anti-protons the systematic uncertainties of $\ptracksys$ and $\pbartracksys$ are obtained, respectively, and a total systematic uncertainty of $\ppbartracksys$ is assigned.
Similar to the tracking uncertainty, the systematic uncertainty for the PID is studied in bins of the momentum and $\cos\theta$.
The corresponding weights are also determined by studying the decay channel $\JPsi\to\pi^+\pi^-\p\pbar$.
For the PID of the proton and anti-proton, the systematic uncertainties are $\ppidsys$ and $\pbarpidsys$, and the total systematic uncertainty of $\ppbarpidsys$ is assigned.

The difference  in the reconstruction of photons is studied with the decay channel $\JPsi\to\gamma\mu\mu$.
The resulting systematic uncertainty is $\goodphotonsys$ for each photon from the $\piz$.
For the $\Sigp$ reconstruction no additional requirement is applied and therefore no systematic uncertainty is assigned.
The $\KS\to\pip\pim$ reconstruction uncertainty is obtained by studying the difference in dependence on the $\KS$ momentum of the decay channels  $\JPsi\to\KS K^\pm\pi^\mp$ and $\JPsi\to\phi\KS K^\pm\pi^\mp$.
By averaging the reconstruction efficiencies of data and MC simulation the systematic uncertainty of $\KSrecsys$ is obtained. For the determination of the systematic uncertainty of the requirement on the minimum angle between a photon and the nearest track $\Delta \alpha$, the requirement is varied by $\pm30\%$. The maximum deviation from the scenario with the nominal requirement is $\danglesys$ and taken as the systematic uncertainty.

The systematic uncertainty on the branching fraction due to the $\Lambda$ and the $\eta$ veto is determined by varying the requirements by $\sim\pm\frac{1}{2}\sigma$ of the width $\sigma$ of the corresponding resonance peak.
For the $\Lambda$ veto no systematic deviation is observed.
For the $\eta$ veto the systematic uncertainty is $\etavetosys$.
It is necessary to correct the helix parameters of the simulated tracks to match the $\chi^2_{\rm kin}$ distribution of the kinematic fit between data and MC simulation.
The difference of the branching fraction with and without this correction is determined to be $\kinsys$ which is taken as the systematic uncertainty. The statistical uncertainty of the number of QED background events $N^{3097}_{\rm QED}$ is propagated as a systematic uncertainty which is $\QEDsys$.

The $M_{\pi^+\pi^-}$ fit procedure depends on three values whose uncertainties have to be taken into account:
the fit range, the definition of the signal range, and the fit model of the background distribution.
The uncertainties of the signal region and  fit range are obtained by changing the size of the corresponding windows by $\pm\num{10}\%$ and remeasuring the branching fraction.
The systematic uncertainty for the signal region is found to be negligible, and for the fit range it is $\FitRangesys$.
For the background description the order of the polynomial is increased and decreased by one.
This yields an uncertainty of $\BKGmodelsys$.

To determine the uncertainty of the reconstruction efficiency due to the amplitude model, the parameters of the amplitude model are varied 1000 times according to the covariance matrix, and the efficiency is reevaluated.
The RMS of the resulting efficiency distribution ($\effmodelsys$) is taken as systematic uncertainty.
The statistical uncertainty of the efficiency is $\effstatsys$ and treated as the systematic uncertainty for the branching fraction measurement.

For the external parameters such as the number of $\JPsi$ events and the branching fractions of the intermediate particles, namely the $\mathcal{B}(\KS\to\pip\pim)$, $\mathcal{B}(\Sigp\to\p\piz)$, and $\mathcal{B}(\piz\to\gamma\gamma)$, error propagation is used.
For $N_{\JPsi}$ this results in \num{0.55}\% \cite{BESIII:2012pbg}, for $\mathcal{B}(\KS\to\pip\pim)$ in \num{0.07}\%, for $\mathcal{B}(\Sigp\to\p\piz)$ in \num{0.58}\%, and for $\mathcal{B}(\piz\to\gamma\gamma)$ in \num{0.03}\% \cite{pdg}.

The total systematic uncertainty is calculated by summing all uncertainties quadratically and taking the square root, resulting in $\totalsys$. The systematic uncertainty corresponding only to the external sources is $0.73\%$

\section{\label{sec::SUMMARY}Summary}

By analyzing $(10\,087\pm44)\times10^{6}$ $\JPsi$ events taken with the BESIII detector, we report the first observation of the decay channels $\SigDecay$ and $\SigbarDecay$.
The branching fractions of these decays are determined to be:
\begin{eqnarray*}
&\BF(\SigDecay)=\BFresultstsySigma \\
&\BF(\SigbarDecay)=\BFresultstsySigmabar
\end{eqnarray*}
\noindent The first uncertainty is statistical and the second systematic. Both results are in good agreement. No difference between the charge conjugate decays is observed. The result of both decays combined is
\begin{eqnarray*}
\BF(\SigDecay+c.c.)=\BFresultstsy,
\end{eqnarray*}
where the first uncertainty is statistical and the second systematic. 
The determined branching fraction is in good agreement with the result of the isospin partner $\JPsi\to pK^-\bar\Sigma^0+c.c.$ measured with the MARKII experiment \cite{PhysRevD29804}.

\section*{Acknowledgments}
\input{Acknowledgements}
\FloatBarrier

\bibliography{Bibliography}

\end{document}

%% file: Results.tex
\newcommand{\BFresultvalue}{\ensuremath{\num{2.725}\pm\num{0.009}}} 
\newcommand{\BFresultstsy}{\ensuremath{\left( \BFresultvalue\pm\num{0.050}\right)\times10^{-4}}} 

\newcommand{\BFresultSigmaValue}{\ensuremath{\num{1.361}\pm\num{0.006}}} 
\newcommand{\BFresultSigmabarValue}{\ensuremath{\num{1.352}\pm\num{0.006}}} 

\newcommand{\BFresultSigma}{\ensuremath{\left( \BFresultSigmaValue \right)\times10^{-4}}}
\newcommand{\BFresultSigmabar}{\ensuremath{\left( \BFresultSigmabarValue \right)\times10^{-4}}}

\newcommand{\BFresultstsySigma}{\ensuremath{\left( \BFresultSigmaValue \pm\num{0.025}\right)\times10^{-4}}}
\newcommand{\BFresultstsySigmabar}{\ensuremath{\left( \BFresultSigmabarValue \pm\num{0.025}\right)\times10^{-4}}}

\newcommand{\BFresultstsyPresentation}{\ensuremath{\left( \BFresultvalue_\text{stat.}\pm\num{0.052}_\text{sys.}\right)\times10^{-4}}}
\newcommand{\BFresultst}{\ensuremath{\left( \BFresultvalue\right)\times10^{-4}}}
\newcommand{\BFpkplusSigma}{\ensuremath{\left(\num{2.9}\pm\num{0.8}\right)\times10^{-4}}}

\newcommand{\NQED}{\ensuremath{ \num{ 270+-70 } } }

\newcommand{\Nsigvalue}{\ensuremath{\num{1.204+-0.004}}}
\newcommand{\Nsigvalueplain}{\ensuremath{\num{120400+-400}}}
\newcommand{\Nsig}{\ensuremath{\left(\Nsigvalue\right)\times10^5}}

\newcommand{\Nsigvaluecorrected}{\ensuremath{\num{1.201+-0.004}}}
\newcommand{\Nsigvaluecorrectedplain}{\ensuremath{\num{120100+-400}}}
\newcommand{\Nsigcorrected}{\ensuremath{\left(\Nsigvalue\right)\times 10^5}}

\newcommand{\effrecvalue}{\ensuremath{\num{12.450+-0.018}}} 
\newcommand{\effrec}{\ensuremath{\left(\effrecvalue\right)\%}}


\newcommand{\NJpsi}{\ensuremath{(\num{10086.6+-43.7})\cdot10^6}}


\newcommand{\ptracksys}{\ensuremath{\num{0.22}\%}} 
\newcommand{\pbartracksys}{\ensuremath{\num{0.33}\%}}
\newcommand{\ppbartracksys}{\ensuremath{\num{0.55}\%}}

\newcommand{\kinsys}{\ensuremath{\num{0.57}\%}} 
\newcommand{\etavetosys}{\ensuremath{\num{0.07}\%}}

\newcommand{\effmodelsys}{\ensuremath{\num{0.19}\%}} 
\newcommand{\effstatsys}{\ensuremath{\num{0.15}\%}} 

\newcommand{\BKGmodelsys}{\ensuremath{\num{0.27}\%}}
\newcommand{\FitRangesys}{\ensuremath{\num{0.13}\%}} 

\newcommand{\danglesys}{\ensuremath{\num{0.25}\%}}

\newcommand{\ppidsys}{\ensuremath{\num{0.21}\%}}
\newcommand{\pbarpidsys}{\ensuremath{\num{0.14}\%}}
\newcommand{\ppbarpidsys}{\ensuremath{\num{0.35}\%}}

\newcommand{\goodphotonsys}{\ensuremath{\num{0.20}\%}}
\newcommand{\totalphotonsys}{\ensuremath{\num{0.40}\%}}

\newcommand{\KSrecsys}{\ensuremath{\num{1.33}\%}}
\newcommand{\QEDsys}{\ensuremath{\num{0.06}\%}}

\newcommand{\totalsys}{\ensuremath{\num{1.85}\%}} 

%% file: AuthorList.tex
\author{
\begin{small}
    \begin{center}
M.~Ablikim$^{1}$, M.~N.~Achasov$^{13,b}$, P.~Adlarson$^{75}$, X.~C.~Ai$^{81}$, R.~Aliberti$^{36}$, A.~Amoroso$^{74A,74C}$, M.~R.~An$^{40}$, Q.~An$^{71,58}$, Y.~Bai$^{57}$, O.~Bakina$^{37}$
% ...
\\
\vspace{0.2cm}
(BESIII Collaboration)\\
\vspace{0.2cm} {\it
$^{1}$ Institute of High Energy Physics, Beijing 100049, People's Republic of China\\
% ...
$^{13}$ G.I. Budker Institute of Nuclear Physics SB RAS (BINP), Novosibirsk 630090, Russia\\
% ...
$^{36}$ Johannes Gutenberg University of Mainz, Johann-Joachim-Becher-Weg 45, D-55099 Mainz, Germany\\
$^{37}$ Joint Institute for Nuclear Research, 141980 Dubna, Moscow region, Russia\\
% ...
$^{40}$ Liaoning Normal University, Dalian 116029, People's Republic of China\\
% ...
$^{57}$ Southeast University, Nanjing 211100, People's Republic of China\\
$^{58}$ State Key Laboratory of Particle Detection and Electronics, Beijing 100049, Hefei 230026, People's Republic of China\\
% ...
$^{71}$ University of Science and Technology of China, Hefei 230026, People's Republic of China\\
% ...
$^{74}$ University of Turin and INFN, (A)University of Turin, I-10125, Turin, Italy; (B)University of Eastern Piedmont, I-15121, Alessandria, Italy; (C)INFN, I-10125, Turin, Italy\\
$^{75}$ Uppsala University, Box 516, SE-75120 Uppsala, Sweden\\
% ...
$^{81}$ Zhengzhou University, Zhengzhou 450001, People's Republic of China\\
\vspace{0.2cm}
% ...
$^{b}$ Also at the Novosibirsk State University, Novosibirsk, 630090, Russia\\
% ...
    }\end{center}
\end{small}
}

%% file: Acknowledgements.tex
The BESIII Collaboration thanks the staff of BEPCII and the IHEP computing center for their strong support. This work is supported in part by National Key R\&D Program of China under Contracts Nos. 2020YFA0406300, 2020YFA0406400; National Natural Science Foundation of China (NSFC) under Contracts Nos. 11635010, 11735014, 11835012, 11935015, 11935016, 11935018, 11961141012, 12022510, 12025502, 12035009, 12035013, 12061131003, 12192260, 12192261, 12192262, 12192263, 12192264, 12192265; the Chinese Academy of Sciences (CAS) Large-Scale Scientific Facility Program; the CAS Center for Excellence in Particle Physics (CCEPP); Joint Large-Scale Scientific Facility Funds of the NSFC and CAS under Contract No. U1832207; CAS Key Research Program of Frontier Sciences under Contracts Nos. QYZDJ-SSW-SLH003, QYZDJ-SSW-SLH040; 100 Talents Program of CAS; The Institute of Nuclear and Particle Physics (INPAC) and Shanghai Key Laboratory for Particle Physics and Cosmology; ERC under Contract No. 758462; European Union's Horizon 2020 research and innovation programme under Marie Sklodowska-Curie grant agreement under Contract No. 894790; German Research Foundation DFG under Contracts Nos. 443159800, 455635585, Collaborative Research Center CRC 1044, FOR5327, GRK 2149; Istituto Nazionale di Fisica Nucleare, Italy; Ministry of Development of Turkey under Contract No. DPT2006K-120470; National Research Foundation of Korea under Contract No. NRF-2022R1A2C1092335; National Science and Technology fund of Mongolia; National Science Research and Innovation Fund (NSRF) via the Program Management Unit for Human Resources \& Institutional Development, Research and Innovation of Thailand under Contract No. B16F640076; Polish National Science Centre under Contract No. 2019/35/O/ST2/02907; The Royal Society, UK under Contract No. DH160214; The Swedish Research Council; U. S. Department of Energy under Contract No. DE-FG02-05ER41374